\documentclass[epj]{svjour}
\usepackage{graphics}
\begin{document}
\title{On the Importance of Isospin Effects for the Interpretation of Nuclear Collisions}
\titlerunning{Importance of Isospin Effects for the Interpretation of Nuclear Collisions}
%\author{First author\inst{1} \and Second author\inst{2}% etc
\author{0nd\v{r}ej Chv\'{a}la\inst{1} \emph{for the NA49 Collaboration}
\thanks{For a full author list of the NA49 Collaboration see \cite{Alt:2003rn} }
%\authorrunning{0nd\v{r}ej Chv\'{a}la \emph{for the NA49}}
}                     % Do not remove
\institute{Institute of Particle and Nuclear Physics, MFF UK, V Hole\v{s}ovi\v{c}k\'{a}ch 2, Praha 8, Czech Republic}
\date{Received: date / Revised version: date}
\abstract{
It is demonstrated that many aspects of nuclear collisions, as for instance the evolution of 
$\pi^{+}/\pi^{-}$ and $K/\pi$ ratios with $x_F$ and $\sqrt{s}$, are influenced by
isospin effects already present in elementary nucleon--nucleon collisions.
\PACS{
      {PACS-key}{describing text of that key}   \and
      {PACS-key}{describing text of that key}
     } % end of PACS codes
} %end of abstract
\maketitle
\section{Introduction}
\label{intro}
The study of heavy ion collisions made at the SPS and the RHIC attracts wide interest.
However, it becomes clearly apparent that the understanding of elementary nucleon-nucleon 
interactions is crucial for the correct interpretation of the more complex nuclear collisions.

One of the basics ingredients to that problem is the role played by the isospin invariance.
Since neutrons constitute 60\,\% of the nucleons inside a heavy ion nucleus,
and since even the spatial distribution of protons and neutrons is known to 
be different in heavy nuclei \cite{pnratnucl}, the proper evaluation of 
isospin effects in proton and neutron fragmentation is of obvious interest.

The NA49 experiment \cite{RefNA49} was the first to measure the
yields of identified hadrons from neutron fragmentation in the SPS energy range \cite{RefHGF}.
In this article, some consequences of these new measurements for relativistic 
nuclear interactions will be presented.
 
\section{$\pi^{+}/\pi^{-}$ ratios}
\label{sec:1}
\begin{figure}[htb]
\resizebox{0.417\textwidth}{!}{%
\includegraphics{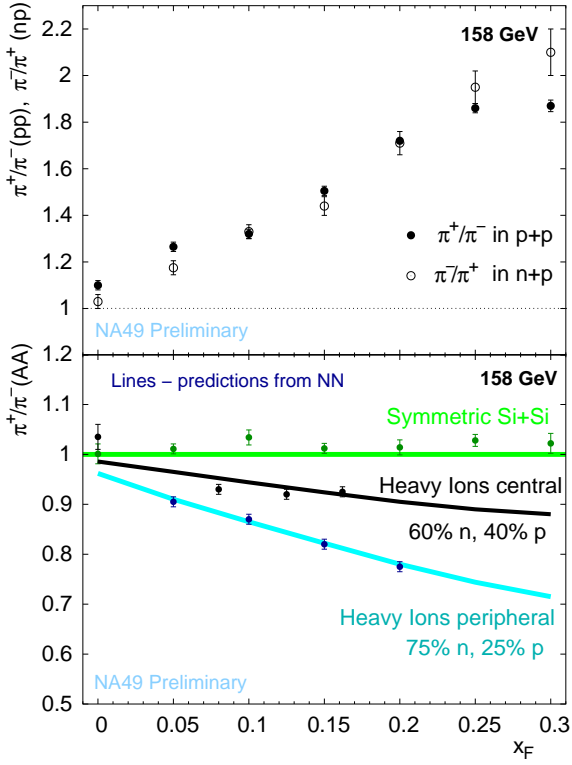}}
\caption{$\pi^{+}/\pi^{-}$ ratios in $pp$, $AA$ and the inverse in $np$ collisions as a function of $x_F$ for different isospin mixtures. NA49 data.}
\label{fig:1}       
\end{figure}

The $\pi^{+}$ and $\pi^{-}$ yields from both the proton and the neutron fragmentation 
have been measured by NA49 \cite{RefHGF}, \cite{RefHGF-isospin}.
As expected from isospin symmetry, the $\pi^{+}$ and the $\pi^{-}$ yields change their place 
when switching from proton to neutron projectiles. 
Consequently, the ratio $\pi^{+}/\pi^{-}$ from protons equals $\pi^{-}/\pi^{+}$ from neutron fragmentation. 
These expectations have been verified for a wide range of $x_F$ and for beam momenta of 40 and 160\,GeV/c. 
See the upper panel in fig. \ref{fig:1} for the latter. For details see \cite{RefHGF-isospin}.

It is known that total and differential pion yields in $AA$ collisions differ only little from a linear 
superposition of nucleon--nucleon collisions according to the number of participant nucleon pairs \cite{ferencQM99}.
It seems therefore reasonable to predict the evolution of the $(\pi^{+}/\pi^{-})^{A}$ 
ratio with the kinematic variables $x_F$ and $\sqrt s$ as a function of $(\pi^{+}/\pi^{-})^{p}$, 
if the detailed behavior of the latter is known:

\begin{equation}
\bigg(\frac{\pi^+}{\pi^-}\bigg)^{A} (x_F, \sqrt s) = 
\frac{f^p \ (\pi^{+}/\pi^{-})^{p} + f^n}{f^p + f^n \ (\pi^{+}/\pi^{-})^{p}} \ (x_f, \sqrt s)
\label{eq:1}
\end{equation}
where $f^p$ and $f^n$ are the relative protonic and neutronic contents of the nuclei -- "isospin mixture", $f^p +f^n = 1$.

\begin{figure}[hbt]
\resizebox{0.428\textwidth}{!}{%
\includegraphics{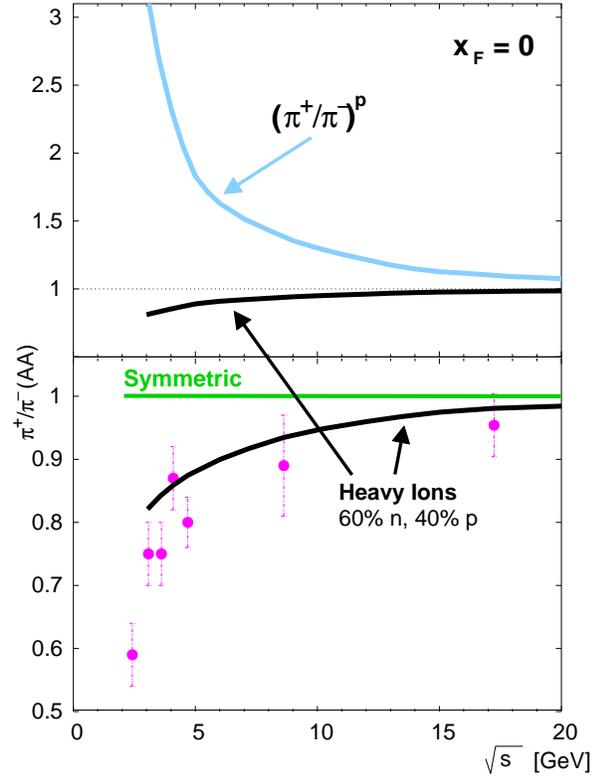}}
\caption{$\pi^{+}/\pi^{-}$ ratios in $pp$ and central $AA$ collisions as a function of $\sqrt s$ at $x_F = 0$}
\label{fig:2}       %
\end{figure}

Evidently, deviations from $(\pi^{+}/\pi^{-})^{A} = 1$ are predicted, growing with $(\pi^{+}/\pi^{-})^{p}$. 
This ratio is both a strong function of $x_F$ and $\sqrt s$, see the upper plots on figures \ref{fig:1} and \ref{fig:2}.

Whereas the measurements of $\pi^{+}/\pi^{-}$ dependence on $x_F$ (fig. \ref{fig:1}, bottom panel) 
in the symmetric $SiSi$ system and in central $PbPb$ collisions follow
the expectation from the above prediction rather closely, the data indicate a substantially higher neutron content in peripheral 
$PbPb$ interactions, as has indeed been established with independent experimental methods \cite{pnratnucl}.

There is a steep dependence of the total and the midrapidity $\pi^+/\pi^-$ ratios on the $\sqrt s$ in $pp$ interactions
from pion production threshold to values close to unity at ISR and RHIC energies. 
The curve represents a parameterization of a large set of existing measurements.
Note the midrapidity ratio in the upper panel in fig. \ref{fig:2} 
(the curve represents a parameterization of a large set of existing measurements), together 
with the prediction for $AA$ using equation \ref{eq:1}.

On the bottom plot, the above prediction is compared with existing measurements in central heavy ion collisions. 
Again the data (see \cite{RefAAsdep} for the data at lower energies) follow the simple superposition picture rather closely. 

\section{$K/\pi$ ratios}
\label{sec:2}
Contrary to pions, the charged kaon yields were measured to be the same from both 
the proton and the neutron projectile fragmentation. This experimental observation has important 
consequences for the $K/\pi$ ratios in $AA$ collisions. This can be exemplified on the basis of double ratios
$(K/\pi)^{A}/(K/\pi)^{p}$ as the kaons drop out from the double ratios.
Simple relations for $K/\pi$ ratios from protons and neutrons, equations \ref{eq:2} and \ref{eq:3}, 
and for arbitrary mixtures of these nucleons, equations \ref{eq:4} and \ref{eq:5}, can therefore be established:

\begin{equation}
\frac{(K^{+}/\pi^{+})^n}{(K^{+}/\pi^{+})^p} = \frac{(\pi^{+})^p}{(\pi^{+})^n} = \bigg(\frac{\pi^{+}}{\pi^{-}}\bigg)^p
\label{eq:2}
\end{equation}
\begin{equation}
\frac{(K^{-}/\pi^{-})^n}{(K^{-}/\pi^{-})^p} = \frac{(\pi^{-})^p}{(\pi^{-})^n} = \bigg(\frac{\pi^{-}}{\pi^{+}}\bigg)^p
\label{eq:3}
\end{equation}

\begin{equation}
\frac{(K^{+}/\pi^{+})^A}{(K^{+}/\pi^{+})^p} = \frac{(\pi^{+}/\pi^-)^p}{f^n + f^p \ (\pi^{+}/\pi^-)^p} 
\label{eq:4}
\end{equation}
\begin{equation}
\frac{(K^{-}/\pi^{-})^A}{(K^{-}/\pi^{-})^p} = \frac{(\pi^{-}/\pi^+)^p}{f^n + f^p \ (\pi^{-}/\pi^+)^p}
\label{eq:5}
\end{equation}

\begin{figure}
\resizebox{0.42\textwidth}{!}{%
\includegraphics{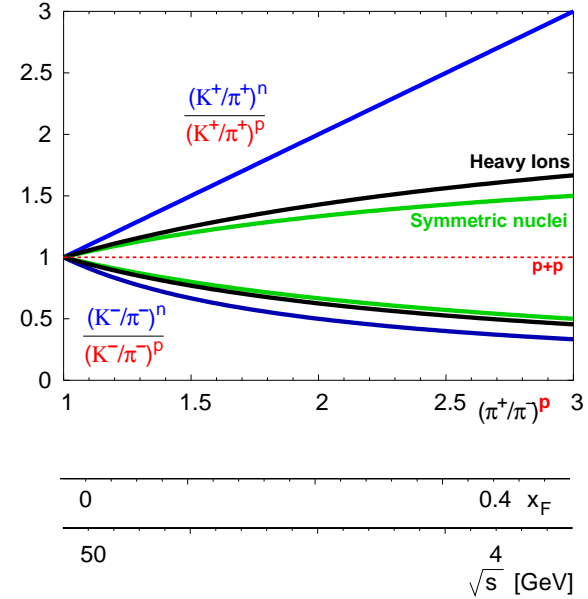}}
\caption{The evolution of $K/\pi$ double ratios with $x_F$ and $\sqrt s$ for different isospin particle fragmentation.}
\label{fig:3}
\end{figure}

Corresponding predictions for $K/\pi$ ratios assuming a linear superposition as used for the $\pi^{+}/\pi^{-}$ ratios
discussed above, are shown in figure \ref{fig:3}. 
Since there is a strong dependence of the $(\pi^{+}/\pi^{-})^p$ ratio on both $\sqrt s$ and on $x_F$, 
we can use the double ratios to make predictions for the evolution of the $K/\pi$ ratios 
in $AA$ with these kinematic variables. Note the scales below the plot in fig. \ref{fig:3}.

The important consequences of the above isospin effects for the interpretation of $K/\pi$ ratios in $AA$ as a function of $x_F$ 
have been demonstrated in \cite{RefHGF}, \cite{RefHGF-isospin}, \cite{strangeness}. It was concluded that the  
enhancements of strange particles in central $pA$ and $AA$ collisions become comparable 
once the isospin effects are corrected for.

\begin{figure}[tb]
\resizebox{0.42\textwidth}{!}{%
\includegraphics{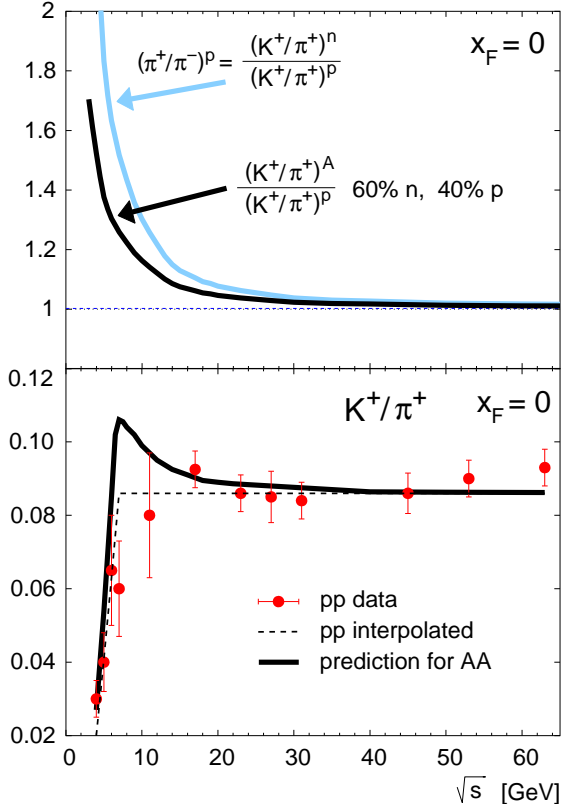}
}
\caption{The evolution of $K/\pi$ double ratios with $\sqrt s$, measured $K^+/\pi^+$ midrapidity ratios in $pp$ and prediction for $AA$.}
\label{fig:4}
\end{figure}

\smallskip
The evolution of $K/\pi$ with $\sqrt s$ is presented in fig. \ref{fig:4}. 
In the upper panel, the $(\pi^{+}/\pi^{-})$ midrapidity ratio from $pp$ collisions is presented.
Note the equation \ref{eq:2}, and the isospin--mixed one according to equation \ref{eq:4}. 
Existing data are plotted in the bottom panel. They were fitted by a flat line with a threshold. 

The isospin correction for positives diverges for decreasing $\sqrt s$. 
This divergence is, however, to be convoluted with the threshold behavior of kaon production. 
Depending on the detailed $\sqrt s$ dependence of the  $K^{+}/\pi^{+}$ ratio in $pp$ collisions, 
the threshold cut--off tends to produce a spike (fig. \ref{fig:4} lower panel) below about 10\,GeV 
in $AA$ collisions. Such non-monotonic behavior is indeed observed in $PbPb$ interactions \cite{Alt:2003rn}. 

For the negatives, the prediction leads to further depletion below the threshold, changing its slope, 
as again observed in $AA$ collisions \cite{Alt:2003rn}.

\begin{figure}[htb]
\resizebox{0.42\textwidth}{!}{%
\includegraphics{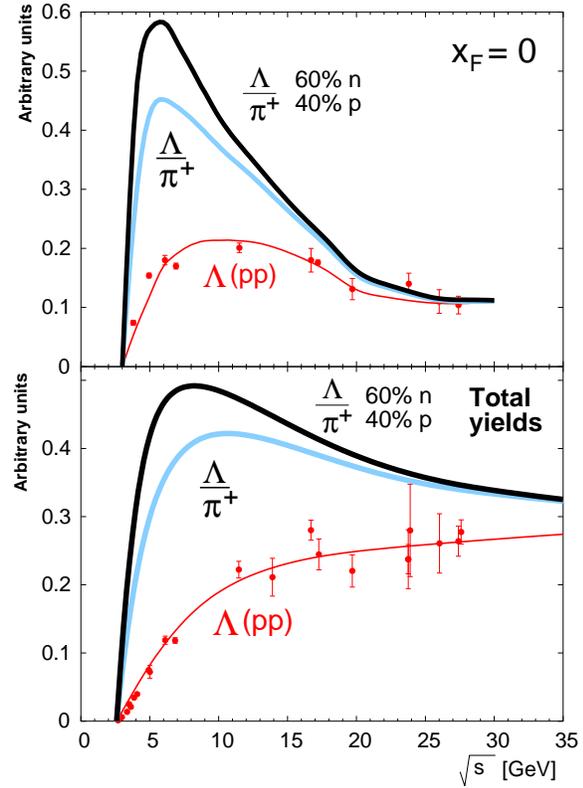}
}
\caption{$\Lambda/\pi$ ratios in $pp$, $\Lambda/\pi$ and prediction for $AA$.}
\label{fig:5}
\end{figure}

A similar phenomenon is predicted for the evolution of $\Lambda/\pi$ ratios, see figure \ref{fig:5}.

\section{Conclusions}
\label{sec:3}
The fragmentation of neutron and proton projectiles into identified secondary hadrons has been measured
at the CERN SPS using $np$ and $pp$ collisions.
Based on these measurements and knowledge of $\pi^+/\pi^-$ ratio in proton-proton collisions,
predictions for $AA$ interactions (assuming that a nuclear collision can be pictured as a sum of independently
fragmenting nucleons) have been formulated.  
The predictions of the evolution of $\pi^+/\pi^-$, $K/\pi$ and $\Lambda/\pi$, 
were found to describe the gross features of the data.

This is especially important for strangeness production in the region of $\sqrt s \ <$ 20\,GeV, where a combination of isospin 
effects and threshold dependencies creates a pronounced, non--monotonic structure.

Improved datasets (in particular for $np$ interactions) are therefore mandatory 
before any conclusions on new phenomena in relativistic heavy ion collisions can be drawn.

\end{document}